\documentclass[10pt]{article}

\usepackage{amsmath}
\usepackage{amssymb}
\usepackage{graphicx}
\usepackage{cite}
\usepackage{color}
\usepackage{setspace} 

\usepackage[labelfont=bf,labelsep=period,justification=raggedright]{caption}

\makeatletter
\renewcommand{\@biblabel}[1]{\quad#1.}
\makeatother

\date{}

\begin{document}
\begin{flushleft}
{\Large
\textbf{Crowding effects in vehicular traffic}
}
\\
Jay Samuel L. Combinido, 
May T. Lim$^\ast$, 
\\
National Institute of Physics, University of the Philippines Diliman, 1101 Quezon City, Philippines
\\
$^\ast$E-mail: may@nip.upd.edu.ph
\end{flushleft}

\doublespacing

\section*{Abstract}
While the impact of crowding on the diffusive transport of molecules within a cell is widely studied in biology, it has thus far been neglected in traffic systems where bulk behavior is the main concern. Here, we study the effects of crowding due to car density and driving fluctuations on the transport of vehicles. Using a microscopic model for traffic, we found that crowding can push car movement from a superballistic down to a subdiffusive state. The transition is also associated with a change in the shape of the probability distribution of positions from negatively-skewed normal to an exponential distribution. Moreover, crowding broadens the distribution of cars' trap times and cluster sizes. At steady state, the subdiffusive state persists only when there is a large variability in car speeds. We further relate our work to prior findings from random walk models of transport in cellular systems. 

\section*{Introduction}
Crowding, termed as macromolecular crowding in biology, is a state where a solution (or a cell membrane) has a high concentration of macromolecules such that a significant volume becomes inaccessible to other molecules \cite{mcrowding}. A crowded environment obstructs the path of diffusing molecules, which in turn reduce the mobility of molecules and leads to anomalous or subdiffusive transport. As such, the mean-square displacement (MSD) of the particles is no longer proportional to time, rather, $\rm{MSD}\propto \mathit{t}^\alpha$ with $\alpha<1$. Experiments through single particle tracking \cite{spt1,spt2} and fractional correlation spectroscopy \cite{anomcyto}, as well as random-walk computer simulations provide evidence that subdiffusion indeed manifests from hindered motion due to immobile and/or mobile obstacles\cite{anom_obs,anom_obs2}. Even in directed cellular transport where particles are normally superdiffusive or ballistic ($\alpha=2$), crowding has been reported to induce sub-ballistic ($\alpha<2$) behavior \cite{diffanddirmotion,enhanceddiff}. Though the effect of crowding on transport have been mostly addressed in the context of cellular systems \cite{proteinfolding,attcrowd}, its effect might be relevant on transport in the field of vehicular traffic ---  a system that is four orders of magnitude larger.

Crowding mechanisms in vehicular traffic follow the same principles as in its microscopic counterpart. While hindered movement in cells is caused by collisions \cite{colcrowd1,colcrowd2}, attractive or repulsive forces \cite{attcrowd}, and trapping \cite{anom_obs,anom_obs2}; car movement is prevented (deceleration) by car-to-car interactions due to obstructions and driving fluctuations. However, unlike in cells, changes in movement due to crowding are not commonly studied in traffic research. Often, fundamental diagrams \cite{traffund1,traffund2,traffund3} are presented from statistical (macroscopic) averages of density, speed and flow to depict traffic properties such as jam formations \cite{jamformation,featjam} and phase transitions \cite{traftrans1,moduturn,traftrans2}. Furthemore, the term ``crowding'' may be confused with two traffic concepts that have been studied in great detail: crowd dynamics (often associated with pedestrians) \cite{walkingbehavior} and traffic jams \cite{trafficreview, trafficreview2, trafficjams}. In (pedestrian) crowd dynamics,  the focus is on determining the local mechanisms (e.g. follow the majority, maintain your distance, or walk with your group \cite{walkingbehavior}) that lead to the system's self-organization. It is aimed at making models with practical applications in safety engineering and crowd control. Traffic jams, where traffic flow is nearly stopped, may stem from the trivial case of car volume exceeding road capacity; but may also stem from car interactions that magnify the effect of random fluctuations in driving speed. The latter case produces the phenomenon of phantom jams \cite{trafficjams}, which occur even when car density seem low, but at car density levels not low enough for car headway to remain unaffected by fluctuations in driving speeds. The goal of this work is to gain a better understanding of traffic by probing the effect of ``vehicular'' crowding (in the biophysical sense) on car movement. Here, we simulate a single-lane freeway traffic using the Nagel-Scheckenberg model \cite{Nasch} to examine the deviation of car motion from its normal ballistic behavior as a result of crowding that is related to road density and driving fluctuations.  We present the scaling exponent of the MSD, $\alpha$, during transient and steady state periods. We also show the time evolution of the probability density, as well as the trap time and cluster size distributions and compare our findings to established results from random walk models. Moreover, we surmise the possible correspondence of the rules of the NaSch model to directed random walks and general particle flow.

\section*{Methods}
\begin{figure}[!ht]
\begin{center}
\includegraphics[scale=1]{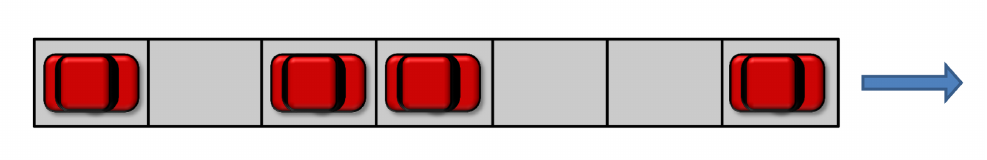}
\end{center}
\caption{
{\bf Road setup} The road is represented by a periodic boundary 1D lattice divided into L=200 cells. Each cell is either empty or occupied by a single car. 
}
\label{fig:setup}
\end{figure}

The Nagel-Schreckenberg (NaSch) model was implemented on a 1D lattice of length $\rm{L}=200$ cells with periodic boundaries (Fig. \ref{fig:setup}). Each cell can be occupied by at most one car. The state of the $i^{\rm{th}}$ car is characterized by its speed $v_i$ = 0,1,...,$v_{\rm{max}}$ and its position $x_i$. Cars are updated in parallel (synchronous update) at each time step $t \leftarrow t + \Delta t~(\Delta t=1)$ according to the following rules:

\vspace{0.4cm}
\noindent\textbf{R1} \textit{Acceleration}. $v_{i} \leftarrow v_{i}+1$ if $v_{i}<v_{\rm{max}}=5~\rm{cells/\Delta \mathit{t}}$.

\noindent\textbf{R2} \textit{Deceleration}. $v_{i} \leftarrow d_{i}/\Delta t$ if $v_{i}\Delta t>d_{i}$ ($d_{i}$ is the number of empty cells in front of a car).

\noindent\textbf{R3} \textit{Slowdown randomization}. $v_{i} \leftarrow v_{i}-1$ with probability $p_i$
  
\noindent\textbf{R4} \textit{Forward movement}. $x_i \leftarrow x_i+v_i\Delta t$

\vspace{0.4cm}
R1 allows cars to accelerate up to a maximum speed $v_{\rm{max}}$. R2 ensures no collision between cars. R3 accounts for driving fluctuations possibly due to non-deterministic acceleration, over-deceleration, or intentional slowing down. A car-dependent randomization probability $(p_i)$ accounts for the randomization diversity among drivers or types of cars being driven, e.g. public vehicles slow down more often than private ones, which plays a crucial role for platoon formations found in realistic freeway traffic. Finally, R4 is the actual motion of the car. Because of the time-parallel update scheme which force car headways to almost always be affected by slowdown randomization, the NaSch model is able to replicate phantom jams \cite{Nasch}.  However, phantom jams are absent in the deterministic case when all $p_i$'s are zero (all cars travel at the speed of the first car) or when all $p_i$'s are one (the $i$th car travels either at its initial speed, or if the car immediately in front of it happens to be moving slower, then at the speed of the preceding car).  Notably, when $v_{max} = 1$, the single-lane NaSch model with no passing is reduced to the well-studied totally asymmetric simple exclusion process (TASEP) model \cite{trafficreview}. 

For each car density $\rho$, a ``perfect'' tracer (a tracked car with randomization probability $p_i=0$) was placed on the first cell of the lattice. The remaining $\rho \rm{L}-1$ cars were placed randomly on the lattice. Each car was assigned an initial speed taken from an empirical speed distribution\cite{moduturn} and was assigned a randomization probability taken from a beta distribution:

\begin{equation}
P(p;a) \propto p^{a-1}\left(1-p\right)^{k-a-1}
\end{equation}
where $0<a<k$. We chose a beta distribution due to its finite range (0 to 1) and versatility: it can represent the range of distributions from a uniform random to a Gaussian distribution with variable skew. The variable $a$ shifts the mean of the distribution $\langle p \rangle = a/k$, while the constant $k$ partially controls the variance given by: $\sigma^2 = a(k-a)/\left[(k^2)(k+1)\right]$. Here, we have arbitrarily chosen the value of $k$ = 10.

Cars states were updated following the rule sequence: R1-R2-R3-R4 during each time step. The simulation was done for different values of $\rho$ (range: [0.1,1.0), interval: 0.05) and $\langle p \rangle$ (range: [0.05,1.0), interval: 0.05). The position of the tracer was recorded every time step and the MSD was obtained through $N$=400 trajectories of different car configurations using the equation: MSD($t$) = $\langle x(t)^2 \rangle$, the average value of tracers' squared displacements at time $t$. 

\section*{Results and Discussion}
\subsection*{Mean-square displacement}
\begin{figure}[!ht]
\begin{center}
\includegraphics[scale=1]{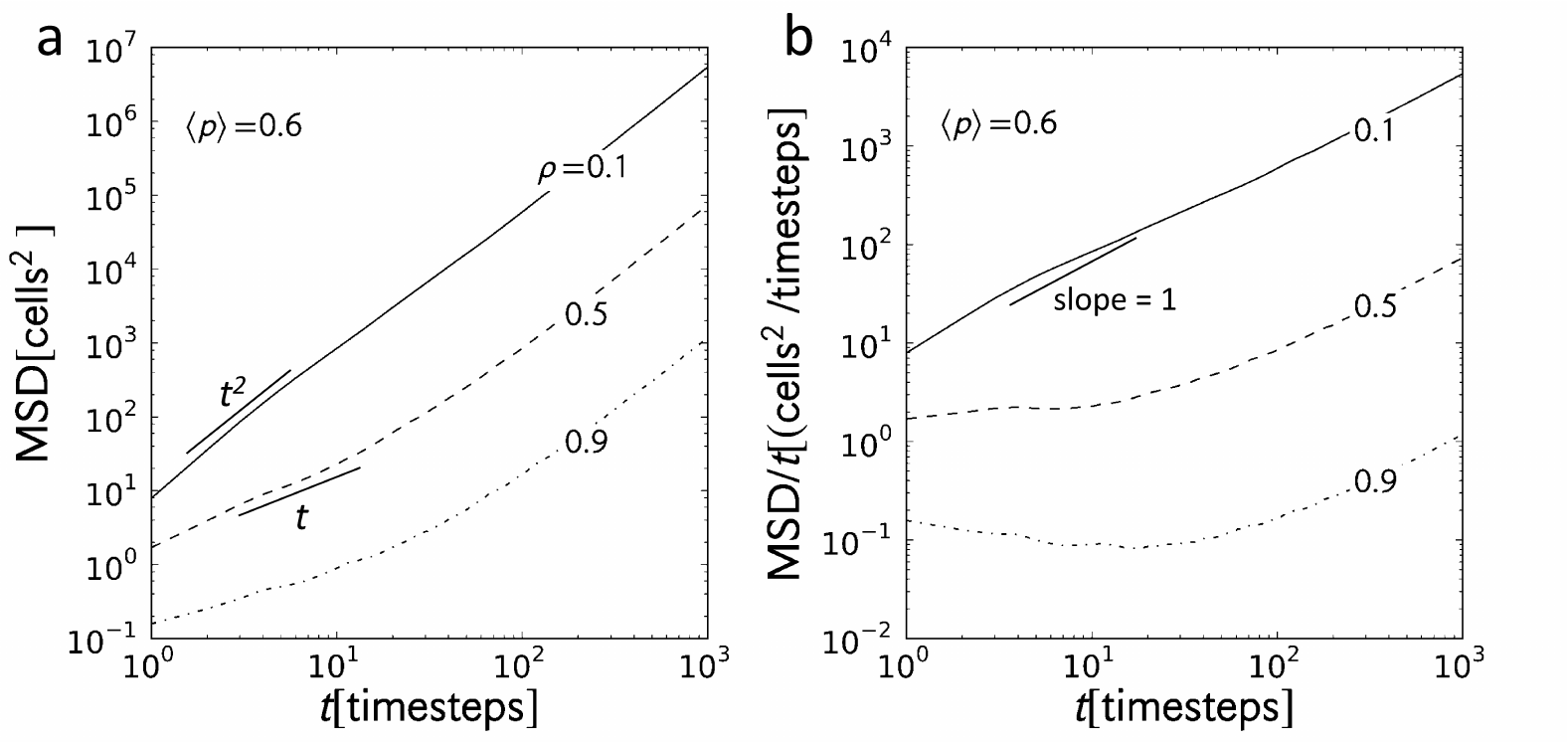}
\end{center}
\caption{
{\bf MSD plots} (a) Mean-square displacement (MSD) as function of time for car densities $\rho=\lbrace 0.1, 0.5, 0.9\rbrace$ at $\langle p \rangle=0.6$. All curves are superdiffusive with $\alpha=2$ at long times but the behavior at short times changes with $\rho$. (b) Plot of the scaled MSD, $\rm{MSD}/\mathit{t}$, reveals that during the transient period, cars can be at subdiffusive states ($\rm{slope} < 0$) at high $\rho$.
}
\label{fig:msdplot}
\end{figure}

Road traffic has two basic parameters which can cause crowding and consequently influence car movement: car density and driving fluctuations. Car density $\rho$ varies inversely with available space for movement. The plot (Fig. \ref{fig:msdplot}a) of the MSD with time for different values of $\rho$ at $\langle p \rangle=0.6$ (high randomization probability) shows two distinct behaviors: a transient behavior that varies with $\rho$ and a steady steady state behavior that is independent of $\rho$. During the transient period, the system self-organizes from its initial random configuration, or relaxes, in such a way that cars redistribute evenly along the stretch of the road and optimize their speeds. The density-dependent transient state implies that the way the system relaxes is affected by the car density $\rho$. To provide a better picture of said dependence, we plotted $\rm{MSD} /\mathit{t}$ against $t$ in Fig. \ref{fig:msdplot}b. Here, diffusion is indicated by a slope of zero. Superdiffusion and subdiffusion yield positive and negative slopes, respectively.  At $\rho=0.1$, the tracers are initially super-ballistic ($\rm{slope}> 1$) which indicate that they accelerate before reaching steady state. Once $\rho$ becomes high, the decrease in available space leads tracers to decelerate sooner. Diffusive transient behavior is observed for $\rho=0.5$. Subdiffusion at $\rho=0.9$ already indicate extreme deceleration. At steady state, however, $\rho$ does not affect the behavior of the cars. In this particular example, cars eventually reach a common ballistic state.

Driving fluctuation or randomization $\langle p \rangle$, the second parameter influencing car movement, prevents acceleration and enhances deceleration of cars. This randomization creates a pseudo-crowded environment that hinders movement regardless of the available space. Just like increased car density, increased $\langle p \rangle$ also causes cars to deviate from their normal ballistic ($\alpha = 2$) motion. 

\begin{figure}[!ht]
\begin{center}
\includegraphics[scale=1]{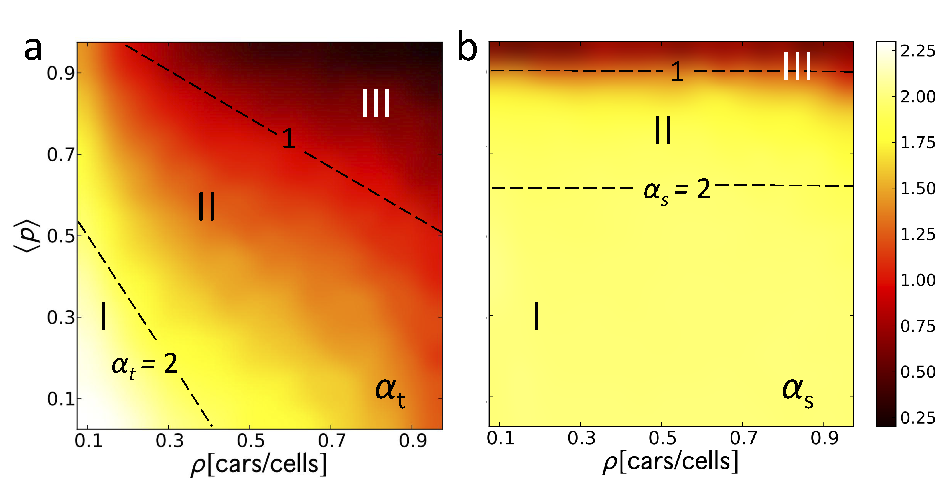}
\end{center}
\caption{
{\bf Transient and steady state scaling exponents} Colormap of the (a) transient and (b) steady state values of the scaling exponent $\alpha_t$ and $\alpha_s$ as a function of car density $\rho$ and the mean randomization probability $\langle p \rangle$. Dashed lines show the approximate super diffusion ($\alpha=2$) and normal diffusion ($\alpha=1$) boundaries. Region I (a) is always superdiffusive ($\alpha\geq2$) indicating a good region for traffic. Region II generally starts with a superdiffusive transient behavior ($1<\alpha<2$) then approach $\alpha = 2$. Lastly, Region III is subdiffusive. The scaling exponent $\alpha_{s}$ is independent of $\rho$.
}
\label{fig:alphamap}
\end{figure}

The joint effects of the two parameters $\rho$ and $\langle p \rangle$ on the the transient and steady state scaling exponents $\alpha_t$ and $\alpha_s$ (Fig. \ref{fig:alphamap}) were calculated using a linear least squares method to determine $\alpha$ in the relation: $\log$(MSD) $\propto \alpha\log(t)$ (as in Fig. \ref{fig:msdplot}). In general, an increase in $\langle p \rangle$ pulls the behavior away from ballistic movement towards the subdiffusive state. At early times (i.e. during transient state, Fig. \ref{fig:alphamap}a), cars approach the subdiffusive state at a lower randomization $\langle p \rangle$ as the car density $\rho$ increases. But after a long time (steady state), the dependence on car density becomes insignificant, and subdiffusion depends largely on $\langle p \rangle$ -- with the extreme case of persistent subdiffusion (Fig. \ref{fig:alphamap}b) at large values of $\langle p \rangle$. We can divide the colormaps into three regions: I) superballistic ($\alpha_{t,s}>2$); II) sub-ballistic--superdiffusive ($1<\alpha_{t,s}<2$); and III) subdiffusive ($\alpha_{t,s}<1$) regions. Dashed lines indicate the $\alpha_{t,s}=2$ and $\alpha_{t,s}=1$ boundaries. In the transient period (Fig. \ref{fig:alphamap}a), a desirable situation for highway traffic occurs when both car density $\rho$ and $\langle p\rangle$ are small (Region I). In this regime, tracers accelerate during the transient period. Region II is mostly sub-ballistic wherein cars slightly decelerate. Region III is the subdiffusive region where rapid deceleration occurs. Moreover, in regions of equal $\alpha_t$, say the $\alpha_t=2$ line, the increase in one parameter  must be compensated by a decrease in the other: when driving fluctuations become considerable, minimal interaction between the tracer and other cars is required to avoid decelerations (hindered movement) and ensure continuous forward motion; and conversely, when the density becomes high, driving fluctuations must be minimized (small $\langle p \rangle$) to sustain movement. At steady state (Fig. \ref{fig:alphamap}b), however, the dynamics changes. Here, the scaling exponent becomes independent of $\rho$ and is only affected by $\langle p \rangle$. If $\langle p \rangle$ is kept below 0.6, tracers will always reach ballistic motion. At $\langle p \rangle >0.6$, tracers begin to move sub-ballistically and at $\langle p \rangle>0.9$ tracers become subdiffusive. In the last case, tracers are already trapped in platoons led by the cars with a high randomization probability resulting to hindered movement thereby causing an overall subdiffusive behavior. These platoon fronts leave large spaces unoccupied producing several local jams. 

\subsection*{Time evolution of the probability density}
\begin{figure}[!ht]
\begin{center}
\includegraphics[scale=1]{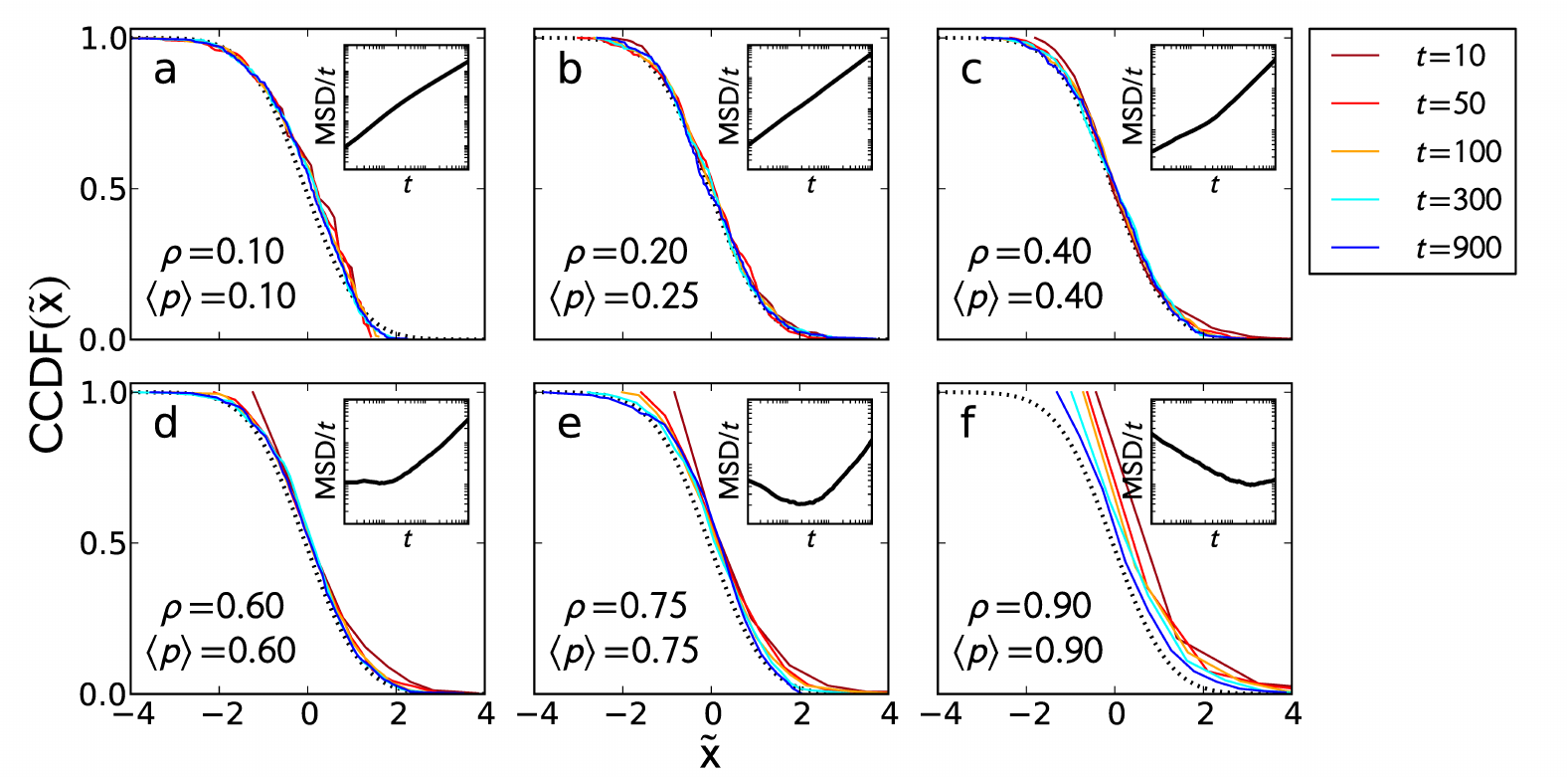}
\end{center}
\caption{
{\bf Time evolution of $\rm{CCDF}(\mathit{\tilde{x}})$}. Time evolution of $\rm{CCDF}(\mathit{\tilde{x}})$ for different car transient behaviors (from superdiffusion [a] to subdiffusion [f]) for $t$ = 10, 50, 100, 300, 900 (dark red to blue). The dotted (black) line is the CCDF of a standard normal distribution. Superballistic cars in (a)  have a CCDF of a negatively skewed normal distribution which can be attributed to the initial acceleration of the cars. Ballistic cars as in (b) fits the standard normal CCDF. Cars with an initial sub-ballistic motion as in (c-f) begin with an exponential distribution then slowly approach standard normal as the movement of cars become ballistic. \textit{Inset}: Plots for the $\rm{MSD}/\mathit{t}$ vs. time \textit{t}. A slope $>0$  indicates superdiffusion; $=0$ is diffusion; and $<0$ is subdiffusion. Ballistic movement is superdiffusive with $\rm{MSD}/\mathit{t}$ vs. \textit{t} slope = 1.
}
\label{fig:probdist}
\end{figure}

Aside from the MSD, the probability distribution of positions $\mathrm{P}(x)$ of the tracers was used to describe the movement of cars. In Fig. \ref{fig:probdist}, we plot the rescaled complementary cumulative density function:\footnote{It is often useful to plot the CCDF rather than the probability distribution  to avoid binning issues and for better data visualization. If one wants to fit a function of a certain probability distribution, taking its CCDF (or alternatively the CDF), fitting it with a function, and taking the derivative will generally do the trick.}
\begin{equation}
\mathrm{CCDF}(\tilde x,t) = 1 - \displaystyle\sum_{\tilde{x_i}<\tilde{x}}\mathrm{P}(X=\tilde {x_i},t)
\end{equation}
of car position z-scores $\tilde x \equiv (x-\langle x \rangle)/\sigma_x$ at different times for representative values of $\rho$ and $\langle p \rangle$ from which superballistic ($\rho=0.10,~\langle p \rangle=0.10;\rm{measured}~\alpha_t=2.3$) down to subdiffusive ($\rho=0.90,~\langle p \rangle=0.90;\rm{measured}~\alpha_t=0.39$) transient periods were observed (see insets). By using the z-score, we removed the position mean and standard deviation information and focused on the time evolution of the shape of the distribution. We juxtaposed the CCDF of a standard normal distribution for comparison. Visually, we observe that at small values of $\rho$ and $\langle p \rangle$, the distributions approximate a normal distribution (Fig. \ref{fig:probdist}). As the value of $\rho$ and $\langle p \rangle$ increases, the distributions become exponentially distributed during a brief transient period and then approach the normal distribution at steady state.

\begin{figure}[!ht]
\begin{center}
\includegraphics[scale=1]{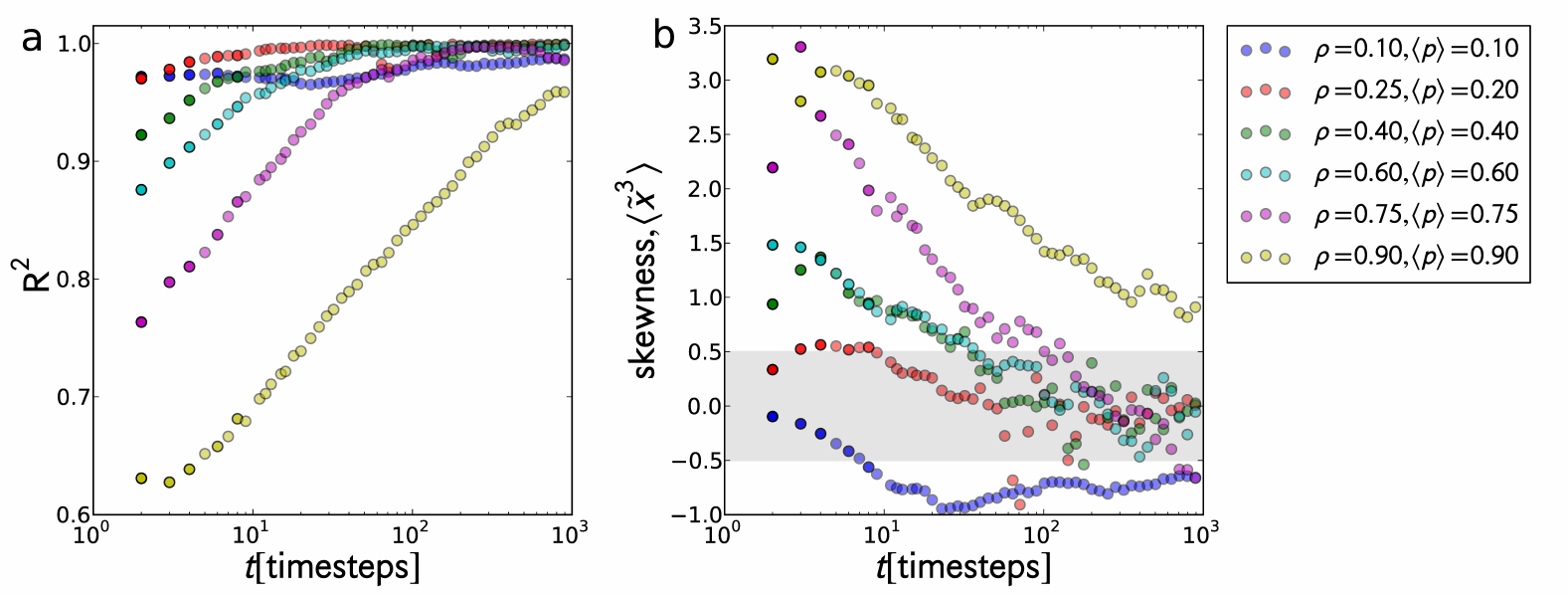}
\end{center}
\caption{
{\bf $\rm{R}^2$ values from Q-Q plots and skewness} The Q-Q plots were calculated from the probability density for all times for $\rho=\langle p \rangle=\lbrace0.1, 0.25, 0.4, 0.6, 0.75, 0.9\rbrace$ to test for normality: (a) $\rm{R}^2$ values were calculated to quantify the closeness of the probability densities to that of a normal distribution; (b) skewness $\langle \tilde{x}^3\rangle$ of the probability densities as a function of time. Values of skewness falling within $|\langle \tilde{x}^3 \rangle|<0.5$ are within the accepted values for a normal distribution. High $\rm{R}^2$ and close to zero $\langle \tilde{x}^3 \rangle$ indicate a normal distribution.
}
\label{fig:r2skew}
\end{figure}

To gain more insight, we tested the obtained z-score distributions for normality by calculating their skewness $\langle \tilde{x}^3 \rangle$, and compared them with a normal distribution using quantile-quantile (Q-Q) plots. The $\rm{R}^2$ values were calculated to quantify the goodness of fit. In Fig. \ref{fig:r2skew} we show the changes on the value of $\rm{R}^2$ (Fig. \ref{fig:r2skew}a) and skewness (Fig. \ref{fig:r2skew}b) as a function of time for various $\rho$ and $\langle p \rangle$. We set the range $|\langle\tilde{x}^3\rangle|<0.5$ (gray region in Fig. \ref{fig:r2skew}b) to be the accepted values for a normal distribution. In the superdiffusive transient case ($\rho=0.10,~\langle p \rangle=0.10$), the distribution is non-normal with $\rm{R}^2<0.99$ and negatively skewed (mass of the distribution is concentrated on the right) which is due to the initial acceleration of the cars. The skewness, however, is hardly recognized (see Fig. \ref{fig:probdist}a) and takes a maximum negative value of approximately $-1.0$ at the transient period and roughly $-0.6$ at steady state. Both values are close to the accepted values for a normal distribution, and the variation can be attributed to the random initial road configuration which do not always guarantee that cars can have an initial acceleration. When tracers move at constant speed all throughout (e.g. $\rho=0.20,~\langle p \rangle=0.25$; Fig. \ref{fig:r2skew}b - red curve), the probability density is most likely, but not necessarily, a normal distribution. Higher values of $\rho$ and $\langle p \rangle$ induce an exponential start of the probability distribution as indicated by the positive skewness and an $\rm{R}^2 < 0.99$ -- tracers have a very small probability of reaching large distances and will likely not even move at all. The trapping of tracers for long periods readily translates, on the average, to a decelerating (sub-ballistic) motion. Eventually, once steady state is reached, the probability density becomes normally distributed. At $\rho=\langle p \rangle = 0.9$ where tracers are always subdiffusive, the exponential distribution is retained for an extended period.

\subsection*{Jam formation}
\begin{figure}[!ht]
\begin{center}
\includegraphics[scale=1]{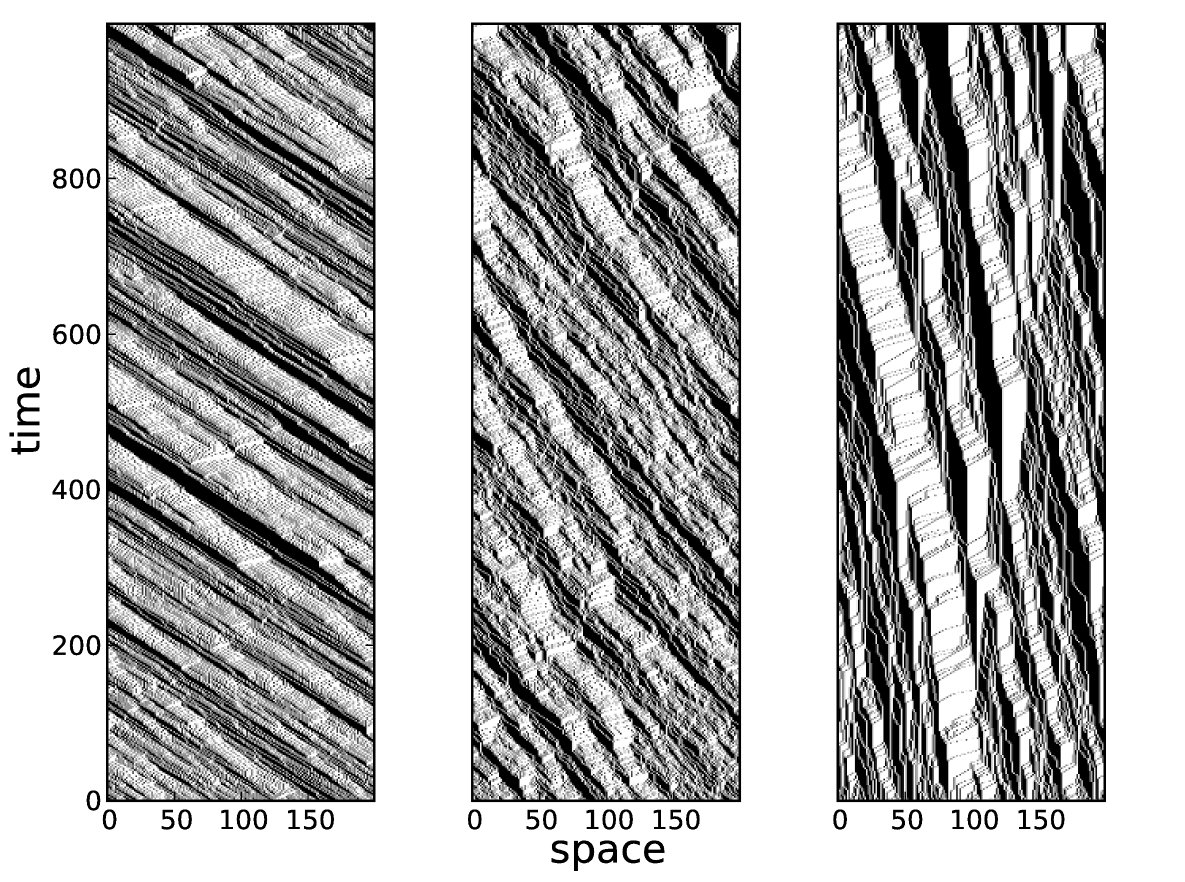}
\end{center}
\caption{
{\bf Spatiotemporal diagrams for varying $\langle p \rangle $} Spatiotemporal diagrams for a constant $\rho=0.5$ at (from left to right) $\langle p \rangle=0.1,0.45,0.80$. The increase in $\langle p \rangle$ produces fewer but larger clusters (black regions) as well as larger unoccupied space.
}
\label{fig:roadsnapshot}
\end{figure}

The decelerations induced by vehicular crowding most often lead to the formation of local jams. In real traffic, jamming is known to occur when the car density becomes appreciable. Larger jams that can sometimes extend to hundreds of meters can form when the density saturates the road capacity. Aside from the density, the randomization also causes cars to form clusters due to platooning effects. Spatiotemporal diagrams (Fig. \ref{fig:roadsnapshot}) show that for high values of randomization probability $\langle p \rangle$, cars leave large spaces unoccupied and thus larger clusters are formed.

\subsubsection*{Cluster size distribution} 
\begin{figure}[!ht]
\begin{center}
\includegraphics[scale=1]{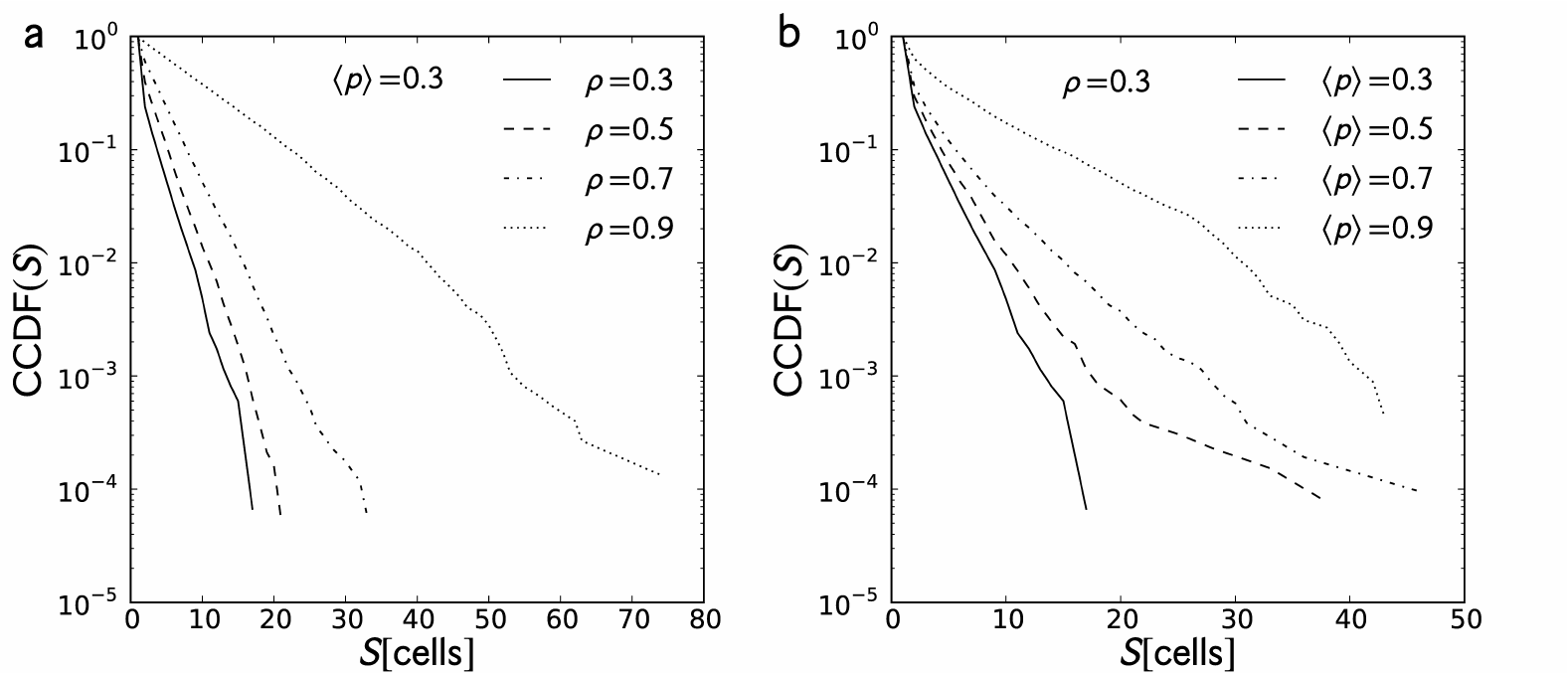}
\end{center}
\caption{
{\bf Cluster size distribution} Cluster size distributions for (a) constant $\langle p \rangle=0.3$ and $\rho = \lbrace 0.3,0.5,0.7,0.9 \rbrace$ and (b) constant $\rho =0.3$ and $\langle p \rangle = \lbrace 0.3,0.5,0.7,0.9 \rbrace$ taken from 400 tracer trajectories. Both distributions are exponential. An increase in either parameters broadens the distributions indicating an increased probability to form large clusters.
}
\label{fig:clustdist}
\end{figure}

\begin{figure}[!ht]
\begin{center}
\includegraphics[scale=1]{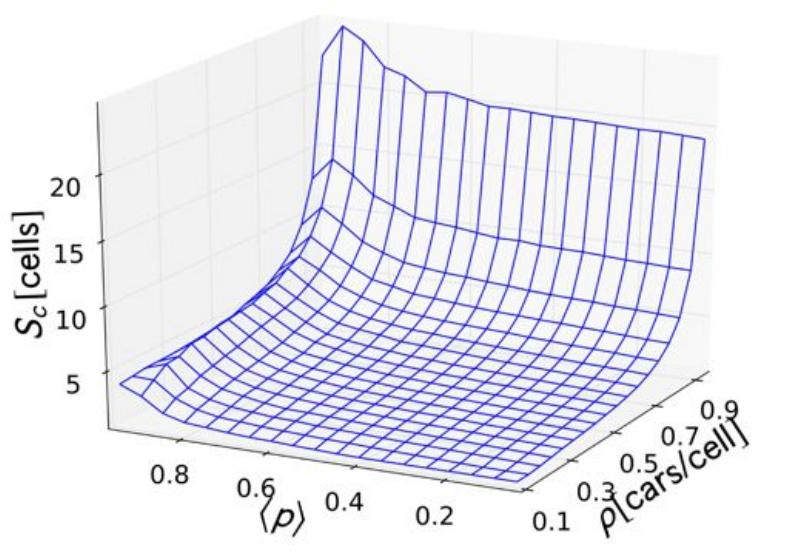}
\end{center}
\caption{
{\bf Cluster size distribution rate parameter} Values of the cluster size distribution rate parameter $S_c$ plotted against $\rho$ and $\langle p \rangle$.
}
\label{fig:csdcconstants}
\end{figure}

To quantify the jam formation, we took the cluster sizes formed by the cars at $t=1000$ from each trial then generated the cluster size distribution from 400 trials. This procedure was done for all $\rho$ and $\langle p \rangle$ values. Fig. \ref{fig:clustdist} shows the CCDF of cluster sizes for representative values of $\rho$ and $\langle p \rangle$. An increase in either parameter, $\rho$ or $\langle p \rangle$ does not change the shape of the distributions but broadens them. The broadening of the distribution with $\rho$ is expected (Fig. \ref{fig:clustdist}a): the increase in the number of cars gives higher chances of forming large clusters. For constant $\rho$ (Fig. \ref{fig:clustdist}b), the broadening of the distribution with increased $\langle p \rangle$ is due to the platoons led by cars with high randomization probability. We can fit an exponential function of the form CCDF($S) \propto \exp\left(-S/S_C\right)$ to the distribution where $S_c$ is a rate parameter that measures the broadening which can be directly obtained using the maximum likelihood estimate (MLE) for exponential distributions\footnote{For exponentially distributed samples $x = \lbrace x_1,x_2,...,x_n\rbrace$ with the probability density described as $P(x) \propto \exp(-\lambda x)$, the maximum likehood estimate of the rate parameter $\lambda$ is given by: $\lambda = 1/\rm{mean}(\mathit{x})$}. A low (high) value of $S_c$ means that the distribution decays fast (slowly) with $S$. The calculated $S_c$ values are plotted against $\rho$ and $\langle p \rangle$ in Fig. \ref{fig:csdcconstants}. Most of the increase is primarily contributed  by $\rho$ because of the increase of the number of cars in the system. Large cluster sizes also form at high $\langle p \rangle$, but are limited by the density; hence the slight increase in $S_c$.

\subsubsection*{Trap time distribution} 
\begin{figure}[!ht]
\begin{center}
\includegraphics[scale=1]{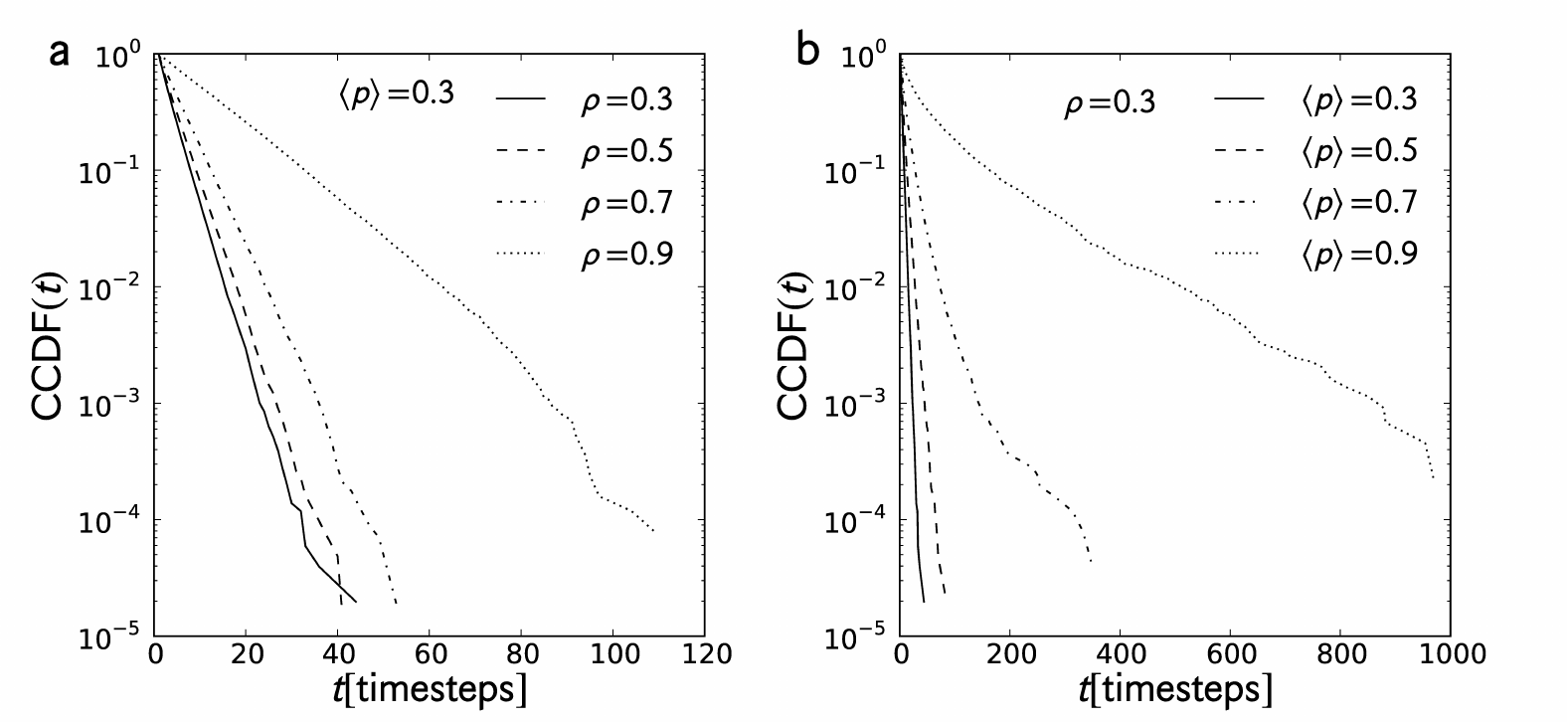}
\end{center}
\caption{
{\bf Trap time distribution} Trap time distributions for (a) constant $\langle p \rangle = \lbrace 0.3,0.5,0.7,0.9 \rbrace$ and $\rho=0.3$ and (b) constant $\langle p \rangle=0.3$ and $\rho = \lbrace 0.3,0.5,0.7,0.9 \rbrace$ taken from 400 tracer trajectories. Both distributions are exponential. An increase in either parameters broadens the distributions indicating an increased probability of tracers being trapped for longer periods.
}
\label{fig:trapdist}
\end{figure}

\begin{figure}[!ht]
\begin{center}
\includegraphics[scale=1]{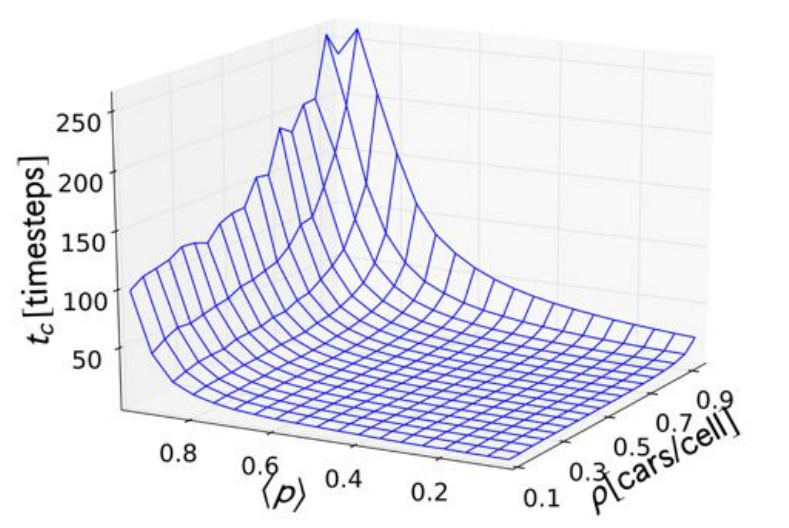}
\end{center}
\caption{
{\bf Trap time distribution rate parameter} Values of the trap time distribution rate parameter $S_c$ plotted against $\rho$ and $\langle p \rangle$.
}
\label{fig:ttdtconstants}
\end{figure}

A tracer's interaction with its environment would normally result to hindered movement, especially in the high-density regime, forcing the tracer to sometimes stop (or be trapped) for a number of time steps. Here, we obtained the trap time distribution from the time intervals that the tracer remains stationary over its entire journey. In Fig. \ref{fig:trapdist}, we show the CCDF of the trap time for constant $\langle p \rangle$ and constant $\rho$, respectively. In both cases, the CCDFs are exponentially distributed thus the probability densities also behave exponentially. Also, the values of $\langle p \rangle$ and $\rho$ do not change the shape of the distributions though an increase in either parameter broadens them. High car density prevents a tracer's movement due to vehicular crowding while with large $\langle p \rangle$, tracers are being trapped by cars with high randomization probabilities. We also calculated the rate parameter $t_c$ and plotted the values in Fig. \ref{fig:ttdtconstants}. We observed that the trap times increase with $\rho$ and $\langle p \rangle$. However, the density $\rho$ has a smaller effect on the trap times compared to $\langle p \rangle$: a high-density situation still allows forward movement, no matter how slow, which causes the tracer to escape the trap. On the other hand, in the high $\langle p \rangle$ regime, tracers are caught behind cars of high randomization probability, which may pause for long periods.

\subsection*{Discussion and Concluding Remarks}
It is interesting to note that these results seem similar to what have been previously observed in random walks. The normal distribution of the CCDF found for ballistic cars can be recovered by solving the diffusion-advection equation \cite{probability} for biased random walks: $P(x,t)\propto \exp[-(x-vt)^2/(4Dt)]$. Our findings for the probability distribution of sub-ballistic cars is likewise observed for subdiffusive random walks \cite{klafter} and can be obtained using the fractional diffusion equation (reviewed by Metzler and Klafter\cite{rev-anodif}). 

Similarly, the exponential distribution of trap times is also seen for Brownian motion when the density of the obstacles increases. Brownian motion simulations also recovers this result when a diffusing particle is surrounded by mobile obstacles [H. Berry and H. Chate (2011), arXiv:1103.2206] and fat-tailed distribution emerges once obstacles become immobile. The resemblance is enough to suggest that the mobility of obstacles is responsible for obtaining such a distribution, whether particles are directional or not. Moving obstacles allow the particle (car) to easily escape traps making it impossible for the particle to be trapped at very long times, hence the fast decay in the trap time distribution. 

Nevertheless, such a correspondence is somewhat expected due to corresponding mechanisms (see Table \ref{tab:corr}) in the simulation rules of the NaSch model, random walks and particle flow. Thus, it is also likely that our findings can be extended to other systems where a movement-maximizing tracer can be embedded into the system. 

\begin{figure}[!ht]
\begin{center}
\includegraphics[scale=1]{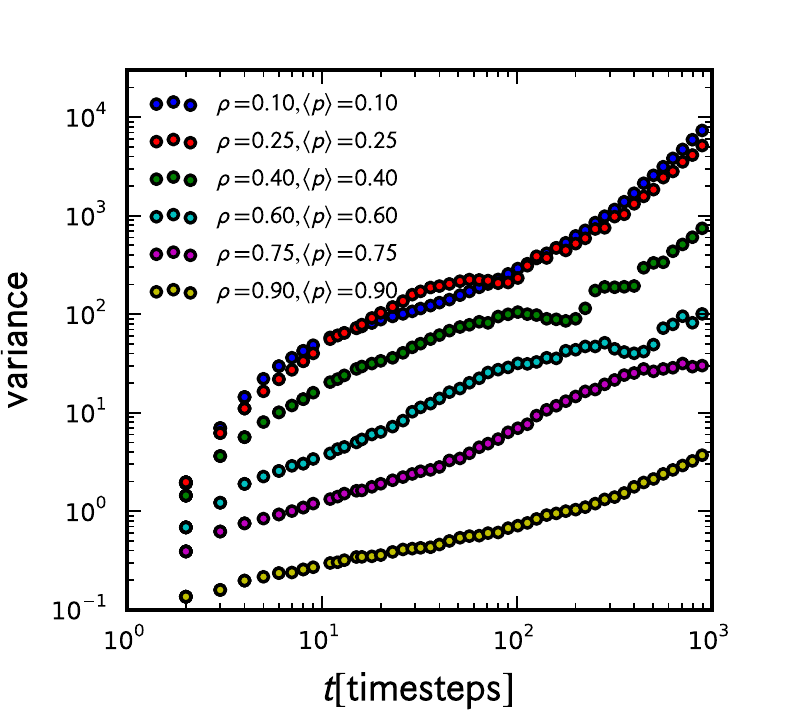}
\end{center}
\caption{
{\bf Spread in the probability density} The variance of the probability density for  $\rho=\langle p \rangle=\lbrace0.1,0.25,0.4,0.6,0.75,0.9\rbrace$ increases with time.
}
\label{fig:variance}
\end{figure}

Despite the similarities, we need to emphasize two dynamical differences that might be easily overlooked when comparing with a biased random walk model. First, the normal distribution cannot be obtained from an ensemble of independent tracers alone. It is also not an artifact of the randomness introduced by R3. Unlike a random walk (whether diffusive or biased) in free space, a tracer in a free road (regardless of its randomization probability $p$ excluding $p=1$ and whether $p$ is taken from any distribution) will continue to accelerate. Thus, for the case of independent tracers, the resulting distribution will always be negatively skewed as in Fig. \ref{fig:probdist}a. Further, the spread (variance) in the distribution (shown in Fig. \ref{fig:variance}) --- which should remain constant over time if the tracers are allowed to move on an empty road --- emerges from the interaction of the tracer with other cars. The decelerations imposed by those interactions are responsible for creating the spread. 
Second, the sub-ballistic behavior, as well as the exponential distribution naturally comes out from the said interaction when  $\rho$ and $\langle p \rangle$ increase at which point hindered movement become more frequent. These properties exhibited by the tracers are thus consequences of vehicular crowding. 

Based on these findings, we can infer that in road segments where car density can inevitably shoot up and/or cannot be easily controlled, especially during rush hour, it is the randomization probability that must be minimized so that cars are able to move at constant rate; i.e., cars moving slowly or those vehicles frequently stopping at specific locations on the road (such as public utility vehicles) must speed up to avoid trailing platoons that can cause a long period of subdiffusion. It is worthwhile to point out that  traffic lights (or any other traffic control device) tend to increase the mean randomization probability of cars - and can be counterproductive when deployed at intersections which do not suffer from traffic jams especially when operating in a synchronized mode, e.g.``green wave'' strategy \cite{greenwave} which tends to reduce the spatial gap variance (or average headway). However, the applicability of ``green wave'' optimization depends on predictable travel times between intersections, and because travel times may not be so predictable with changing car density, self-organizing schemes \cite{gershenson} which adapt to, rather than optimize, traffic are perhaps more suitable for urban traffic.

That crowding gives different observable effects from the point of view of particles during the transient and steady states is an effect that drivers often observe between city traffic and highway traffic. On the other hand, one can also think of interventions which tend to lengthen the travel distance while minimizing driving fluctuations, such as the use of median U-turns in place of left-turns \cite{moduturn}, as attempts to push the behavior of urban traffic to behave similarly to that of highway traffic. Using the empirical equivalent of tracers, i.e. extended floating car data collection \cite{floatingcar} (which can now be mass-deployed via apps deployed on GPS-enabled phones),  a ``crowding" perspective can provide further insights on real traffic over large road networks.


\singlespacing
\section*{Acknowledgments}
One of us (M.T.L.) would like to thank the Abdus Salam International Centre for Theoretical Physics for their hospitality during part of this work. We also thank an anonymous referee for comments on the manuscript. This work is supported by a University of the Philippines System Research Grant.
\bibliography{references}

\begin{thebibliography}{10}
\providecommand{\url}[1]{\texttt{#1}}
\providecommand{\urlprefix}{URL }
\expandafter\ifx\csname urlstyle\endcsname\relax
  \providecommand{\doi}[1]{doi:\discretionary{}{}{}#1}\else
  \providecommand{\doi}{doi:\discretionary{}{}{}\begingroup
  \urlstyle{rm}\Url}\fi
\providecommand{\bibAnnoteFile}[1]{%
  \IfFileExists{#1}{\begin{quotation}\noindent\textsc{Key:} #1\\
  \textsc{Annotation:}\ \input{#1}\end{quotation}}{}}
\providecommand{\bibAnnote}[2]{%
  \begin{quotation}\noindent\textsc{Key:} #1\\
  \textsc{Annotation:}\ #2\end{quotation}}
\providecommand{\eprint}[2][]{\url{#2}}

\bibitem{mcrowding}
Ellis RJ (2001) Macromolecular crowding: an important but neglected aspect of
  the intracellular environment.
\newblock Curr Opin Struct Biol 11: 114-119.
\bibAnnoteFile{mcrowding}

\bibitem{spt1}
Sch\"{u}tz M, Schindler H, Schmidt T (1997) Single-molecule microscopy on model
  membranes reveals anomalous diffusion.
\newblock Biophys J 73: 1073-1080.
\bibAnnoteFile{spt1}

\bibitem{spt2}
Smith P, Morrison I, Wilson K, Fernandez N, Cherry R (1999) Anomalous diffusion
  of major histocompatibility complex class i molecules on hela cell determined
  by single particle tracking.
\newblock Biophys J 76: 3331-3344.
\bibAnnoteFile{spt2}

\bibitem{anomcyto}
Weiss M, Elsner M, Kartberg F, Nilsson T (2004) Anomalous diffusion is a
  measure of cytoplasmic crowding in living cells.
\newblock Biophys J 87: 3518-3524.
\bibAnnoteFile{anomcyto}

\bibitem{anom_obs}
Saxton M (1994) Anomalous diffusion due to obstacles: A monte carlo study.
\newblock Biophys J 66: 394-401.
\bibAnnoteFile{anom_obs}

\bibitem{anom_obs2}
Wedemeier W, Merlitz H, Langowski J (2009) Anomalous diffusion in the presence
  of anomalous obstacles.
\newblock EPL 88: 38004.
\bibAnnoteFile{anom_obs2}

\bibitem{diffanddirmotion}
Caspi A, Granek R, Elbaum M (1998) Diffusion and directed motion in cellular
  transport.
\newblock Phys Rev E 58: 011916.
\bibAnnoteFile{diffanddirmotion}

\bibitem{enhanceddiff}
Caspi A, Granek R, Elbaum M (2000) Enhanced diffusion in active intracellular
  transport.
\newblock Phys Rev Lett 85: 5655-5658.
\bibAnnoteFile{enhanceddiff}

\bibitem{proteinfolding}
Berg B, Ellis RJ, Dobson C (1999) Effects of macromolecular crowding on protein
  folding and aggregation.
\newblock The EMBO Journ 18: 6927-693.
\bibAnnoteFile{proteinfolding}

\bibitem{attcrowd}
Zhou H, Rivas G, Milton A (2008) Macromolecular crowding and confinement:
  biophysical, and potential physiological consequences.
\newblock Annu Rev Biophys 37: 375-397.
\bibAnnoteFile{attcrowd}

\bibitem{colcrowd1}
Wong J, Gierasch L (2010) Macromolecular crowding remodels the energy landscape
  of a protein by favoring a more compact unfolded state.
\newblock J Am Chem Soc 132: 10445-10452.
\bibAnnoteFile{colcrowd1}

\bibitem{colcrowd2}
Luby-Phelps K, Weisiger R (1996) Role of cytoarchitecture in cytoplasmic
  transport.
\newblock Comp Biochem Physiol 115B: 295-306.
\bibAnnoteFile{colcrowd2}

\bibitem{traffund1}
He S, Guan W, Song L (2010) Explaining traffic patterns at on-ramp vicinity by
  a driver perception model in the framework of three-phase traffic theory.
\newblock Physica A 389: 825-836.
\bibAnnoteFile{traffund1}

\bibitem{traffund2}
Tanaka K, Nagatani T, Masukura K (2008) Fundamental diagram in traffic flow of
  mixed vehicles on multi-lane highway.
\newblock Physica A 387: 5583-5596.
\bibAnnoteFile{traffund2}

\bibitem{traffund3}
Ding J, Juang H (2008) A cellular automata model for traffic flow with
  consideration of the inertial driving behavior.
\newblock IJMPC 21: 549-557.
\bibAnnoteFile{traffund3}

\bibitem{jamformation}
Sugiyama Y, Fukui M, Kikuchi M, Hasebe K, Nakayama A, et~al. (2008) Traffic
  jams without bottlenecks--experimental evidence for the physical mechanism of
  the formation of jams.
\newblock New Journal of Physics 10: 033001.
\bibAnnoteFile{jamformation}

\bibitem{featjam}
Kerner B, Klenov S, Hitler A, Rehborn H (2006) Microscopic features of moving
  traffic jams.
\newblock Phys Rev E 73: 046107.
\bibAnnoteFile{featjam}

\bibitem{traftrans1}
BSKerner, Klenov S (2003) Microscopic theory of spatial-temporal congested
  traffic patterns at highway bottlenecks.
\newblock Phys Rev E 68: 036130.
\bibAnnoteFile{traftrans1}

\bibitem{moduturn}
Combinido J, Lim M (2010) Modeling u-turn traffic flow.
\newblock Physica A 389: 3640-3647.
\bibAnnoteFile{moduturn}

\bibitem{traftrans2}
Lee H, Barlovic R, Schreckenberg M, Kim D (2004) Mechanical restriction versus
  human overreaction triggering congested traffic states.
\newblock Phys Rev Lett 92: 238702.
\bibAnnoteFile{traftrans2}

\bibitem{walkingbehavior}
Moussaïd M, Perozo N, Garnier S, Helbing D, Theraulaz G (2010) The walking
  behaviour of pedestrian social groups and its impact on crowd dynamics.
\newblock PLoS One 5: e10047.
\bibAnnoteFile{walkingbehavior}

\bibitem{trafficreview}
Helbing D (2001) Traffic and related self-driven many-particle systems.
\newblock Rev Mod Phys 73: 1067–1141.
\bibAnnoteFile{trafficreview}

\bibitem{trafficreview2}
Nagatani T (2002) The physics of traffic jams.
\newblock Reports on Progress in Physics 65: 1331.
\bibAnnoteFile{trafficreview2}

\bibitem{trafficjams}
Sugiyama Y, Fukui M, Kikuchi M, Hasebe K, Nakayama A, et~al. (2008) Traffic
  jams without bottlenecks—experimental evidence for the physical mechanism
  of the formation of a jam.
\newblock New Journal of Physics 10: 033001.
\bibAnnoteFile{trafficjams}

\bibitem{Nasch}
Nagel K, Schreckenberg M (1992) A cellular automaton model for freeway traffic.
\newblock J Physique I 2: 2221-2229.
\bibAnnoteFile{Nasch}

\bibitem{probability}
Grimmett G, Stirzaker D (2001) Probability and random processes.
\newblock Oxford University Press, 3rd edition.
\bibAnnoteFile{probability}

\bibitem{klafter}
Klafter J, Shlesinger M, Zumofen G, Blumen A (1992) Scale invariance in
  anomalous diffusion.
\newblock Phil Mag B 65: 755-765.
\bibAnnoteFile{klafter}

\bibitem{rev-anodif}
Metzler R, Klafter J (2011) The random walk's guide to anomalous diffusion: a
  fractional dynamics approach.
\newblock Phys Rep 339: 1-77.
\bibAnnoteFile{rev-anodif}

\bibitem{greenwave}
Brockfeld E, Barlovic R, Schadschneider A, Schreckenberg M (2001) Optimizing
  traffic lights in a cellular automaton model for city traffic.
\newblock Phys Rev E 64: 056132.
\bibAnnoteFile{greenwave}

\bibitem{gershenson}
Gershenson C (2005) Self-organizing traffic lights.
\newblock Complex Systems 16: 29-53.
\bibAnnoteFile{gershenson}

\bibitem{floatingcar}
Messelodi S, Modena CM, Zanin M, Natale FGD, Granelli F, et~al. (2009)
  Intelligent extended floating car data collection.
\newblock Expert Systems with Applications 36: 4213 - 4227.
\bibAnnoteFile{floatingcar}

\end{thebibliography}
\newpage
\section*{Figures}
\newpage
\newpage
\newpage
\newpage
\newpage
\newpage
\newpage
\newpage
\newpage
\newpage
\newpage
\section*{Tables}
\begin{table}[!ht]
\caption{
\bf{Correspondence of the NaSch model rules to the dynamics of other systems}}
\begin{tabular}{lccc}
\hline
NaSch model rules & Road traffic & Random walks & Particle flow (blood, etc.)\\
\hline\hline \textbf{R1} & acceleration & direction bias & pressure/external force \\
\textbf{R2} & deceleration & trapping due to & particle collision + \\
 & & mobile obstacles & momentum transfer \\
\textbf{R3} & driving fluctuation & --- & energy dissipation \\
\textbf{R4} & movement & movement & movement\\
\hline
\end{tabular}
\label{tab:corr}
\end{table}

\end{document}